\documentclass[aip,jcp,groupedaddress,reprint,a4paper,amsmath,amssymb]{revtex4-1}
\usepackage[]{bm,epsfig,graphicx}
\usepackage{dcolumn}

\newcommand{\clc}[1]{\multicolumn{1}{c}{#1}}
\newcommand{\be}{\begin{equation}}
\newcommand{\ee}{\end{equation}}

\begin{document}

\title{Rovibrational dynamics of the strontium molecule in the 
A$^1\Sigma_u^+$, c$^3\Pi_u$, and a$^3\Sigma_u^+$ manifold from
state-of-the-art {\em ab initio} calculations}

\author{\sc Wojciech Skomorowski 
} 
\affiliation{\sl Quantum Chemistry Laboratory, Department of Chemistry, University of Warsaw, Pasteura 1,
02-093 Warsaw, Poland}
\author{\sc Filip Paw\l owski}
\address{\sl Physics Institute, Kazimierz Wielki University, pl. Weyssenhoffa 11, 85-072 Bydgoszcz, Poland}
\address{\sl Quantum Chemistry Laboratory, Department of Chemistry, University of Warsaw, Pasteura 1,
02-093 Warsaw, Poland}
\author{\sc Christiane P. Koch}
\affiliation{\sl Theoretische Physik, Universit\"at Kassel,
Heinrich-Plett-Stra{\ss}e 40, 34132 Kassel, Germany}
\author{\sc Robert Moszynski\footnote[1]{Author for correspondence;
e-mail: robert.moszynski@tiger.chem.uw.edu.pl}} 
\affiliation{\sl Quantum Chemistry Laboratory, Department of Chemistry, University of Warsaw, Pasteura 1,
02--093 Warsaw, Poland}

\begin{abstract}
State-of-the-art {\em ab initio} techniques have been applied to compute the
potential energy curves for the electronic states in the 
A$^1\Sigma_u^+$, c$^3\Pi_u$, and a$^3\Sigma_u^+$ manifold
of the strontium dimer, the spin-orbit and nonadiabatic coupling matrix
elements between the states in the manifold, and the electric transition dipole
moment from the ground X$^1\Sigma_g^+$ to the nonrelativistic and relativistic
states in the A+c+a manifold. The potential energy curves and transition moments were 
obtained with the linear response (equation of motion) coupled cluster
method limited to single, double, and linear triple excitations for the
potentials and limited to single and double excitations for the transition
moments. The spin-orbit and nonadiabatic coupling matrix elements were
computed with the multireference configuration interaction method limited to 
single and double excitations. Our results for the nonrelativistic and
relativistic (spin-orbit coupled) potentials deviate substantially  from 
recent {\em ab initio} calculations. The potential energy curve for the
spectroscopically active (1)$0_u^+$ state is in quantitative agreement with
the empirical potential fitted to high-resolution Fourier transform spectra
[A. Stein, H. Kn\"ockel, and E. Tiemann, Eur. Phys. J. D {\bf 64}, 227 (2011)].
The computed {\em ab initio} points were fitted to physically sound analytical
expressions, and used in converged coupled channel calculations of the rovibrational
energy levels in the A+c+a manifold and line strengths for the 
${\rm A}^1\Sigma_u^+\leftarrow {\rm X}^1\Sigma_g^+$ transitions. 
Positions and lifetimes of quasi-bound Feshbach resonances lying above the 
$^1{\rm S_0 + {^3P_1}}$ dissociation limit were also obtained.
Our results reproduce (semi)quantitatively the experimental data observed thus
far. Predictions for on-going and future experiments are also reported.
\end{abstract}
\maketitle

\section{Introduction}
\label{sec1}
In recent years the strontium diatomic molecule, Sr$_2$, has attracted the interest of
theoreticians and experimentalists. Similarly to the calcium dimer,
Sr$_2$ in its ground X$^1\Sigma_g^+$ state does not form
a chemical bond. Indeed, the binding energy of 1081.6 cm$^{-1}$ \cite{Tiemann:10}
is typical for weakly bound complexes rather than for chemically
bound molecules. However, excited states of Sr$_2$ are strongly
bound, and have been observed in many experiments. As a matter of
fact, the strontium molecule was the subject of numerous high-resolution
spectroscopic studies in the gas phase \cite{Liao:80,Gerber:84} and
in rare gas matrices \cite{Andrews:77,Andrews:78}.
The dissociation energy of the ground state 
was first estimated from the RKR inversion of the spectroscopic
data for the ${\rm B}^1\Sigma_u^+\leftarrow {\rm X}^1\Sigma_g^+$ transitions
\cite{Gerber:84}. Recently, a more elaborate study of Sr$_2$ was 
reported \cite{Tiemann:10}, in which measurements of the 
${\rm B}^1\Sigma_u^+\leftarrow {\rm X}^1\Sigma_g^+$ transitions covering 
large parts of the ground state well were recorded.
The spectrum corresponding to the 
${\rm B}^1\Sigma_u^+\leftarrow {\rm X}^1\Sigma_g^+$ transitions
could easily be assigned using standard spectroscopic techniques because
the B state dissociating into ${\rm ^1S+{^1P}}$ atoms
is relatively well isolated, thus not perturbed by any
other electronic state. While the same is true for the ${{\rm A}^\prime}^1\Pi_u$
state reported in Ref. \cite{Tiemann:11}, 
this is not the case for the A$^1\Sigma_u^+$
state dissociating into ${\rm ^1S + {^1D}}$ atoms.
Here the potential energy curve of the A state crosses the curve
of the ${\rm c}^3\Pi_u$ state dissociating into ${\rm ^1S + {^3P}}$ states, and
the corresponding spectrum cannot easily be assigned. 
Experimental investigation of the A state is
limited to Ref.~\cite{Tiemann:11}.  Tiemann and collaborators measured the
spectrum corresponding to the
${\rm A}^1\Sigma_u^+\leftarrow {\rm X}^1\Sigma_g^+$ transition by high-resolution Fourier
transform spectroscopy, and showed that the rovibrational levels of the A 
state get strongly perturbed by the ${\rm c}^3\Pi_u$ state \cite{Tiemann:11}. 
Unfortunately, the amount of the measured data was not sufficient to 
apply a deperturbation procedure that could be used with trust to determine 
the spectroscopic constants of the A$^1\Sigma_u^+$ and c$^3\Pi_u$ states.
In particular, the key information on the spin-orbit coupling between
the A and c states could not be determined from the analysis 
of the spectra, and only an effective potential was obtained.

Most of the {\em ab initio} calculations on the Sr$_2$ molecule
reported in the literature thus far are concerned with the ground state potential energy 
curve \cite{Jones:79,Ballone:91} and the van der Waals constants governing the 
long-range behavior of the ground state potential 
\cite{Stanton:94,Lima:05,Mitroy:2003,Derevianko:2006,Mitroy:2010}. 
To the best of our knowledge only three theoretical papers
considered the excited states of the strontium dimer \cite{Frecon:96,%
Czuchaj:03,Koto:08}. However, in view of the recent findings of Tiemann
and collaborators \cite{Tiemann:11} the quality of these data is
questionable.

It should be stressed that alkaline-earth atoms and molecules are not
only interesting for conventional spectroscopy, but are also intensely
investigated in experiments at ultralow temperatures.
Closed-shell atoms such as alkaline-earth metal atoms are much more 
challenging to cool and trap than open-shell atoms like the alkali atoms.
They do not have magnetic moments in the ground state
that would enable magnetic trapping. Moreover, 
the short lifetime of the first excited $^1$P$_1$ state implies rather
high Doppler temperatures, requiring a dual-stage cooling with
the second stage operating near the $^3{\rm P}_1$ intercombination line.
Despite these challenges, cooling of calcium, strontium, and ytterbium
atoms to micro-Kelvin temperatures has been realized, and Bose-Einstein
condensates of $^{40}$Ca \cite{KraftPRL09},
$^{84}$Sr \cite{StellmerPRL09,EscobarPRL09},
$^{86}$Sr \cite{StellmerPRA10}, $^{88}$Sr \cite{MickelsonPRA10},
$^{170}$Yb \cite{FukuharaPRA07}, and $^{174}$Yb \cite{TakasuPRL03}
have been obtained.

On the other hand, the closed-shell structure of the alkaline-earth
metal atoms leads to very simple  molecular potentials
with low radiative losses and weak coupling to the environment.
This opens new areas of possible applications, such as
manipulation of the scattering properties with low-loss optical
Feshbach resonances \cite{CiuryloPRA05},
high-resolution photoassociation spectroscopy
at the intercombination line \cite{TojoPRL06,ZelevinskyPRL06},
precision measurements to test for a  time variation of the
proton-to-electron mass ratio \cite{ZelevinskyPRL08,ZelevinskyPRA}
and of the fine structure constant \cite{Schwerdtfeger:12},
quantum computation with trapped polar molecules \cite{DeMille:02}, and
ultracold chemistry \cite{Ospelkaus:2010}.

Of particular interest for the present work are the experimental
investigations of the photoassociation spectra near the
intercombination line 
\cite{TojoPRL06,ZelevinskyPRL06,KillianPRA08} and proposed precision 
measurements of fundamental constants
\cite{ZelevinskyPRL08,ZelevinskyPRA}. In particular the latter require 
a precise route for the production of ultracold molecules in
predefined rovibrational states. This, in turn, requires an accurate 
knowledge of the potential energy curves and the various couplings that
may occur in the A$^1\Sigma_u^+$, c$^3\Pi_u$, and a$^3\Sigma_u^+$ 
manifold of electronic states. Also,
the assignment of the photoassociation spectrum requires a
detailed knowledge of the rovibrational levels close to the dissociation
limit. At present, the experimental data on the A state recorded to date
 \cite{Tiemann:11} is far from complete.
Therefore, in the present paper we report a theoretical study of the
spectroscopy of the strontium dimer in the A$^1\Sigma_u^+$,
c$^3\Pi_u$, and a$^3\Sigma_u^+$ manifold by state-of-the-art {\em ab
  initio} methods.  
The plan of this paper is as follows. In sec. \ref{sec2} we describe 
the {\em ab initio} electronic structure and quantum dynamical
calculations. We present the numerical results in sec. \ref{sec3} and 
discuss at length the accuracy of the present results, 
compare with the available experimental data, and report predictions for the
on-going experiments \cite{Zelevinsky:prive}. Finally, in
sec. \ref{sec4} we conclude our paper.


\section{Computational details}
\label{sec2}
\subsection{{\em Ab initio} electronic structure calculations}
\label{sub1}
In the present study we adopt the computational scheme successfully applied
to the ground and excited states of the calcium dimer
\cite{Moszynski:03,Moszynski:05,Moszynski:06,Moszynski:06a,Moszynski:08}, 
magnesium dimer \cite{Koch:11,Koch11a}, (BaRb)$^+$ molecular ion
\cite{Krych:11}, and SrYb heteronuclear molecule \cite{Tomza:11}. 
The potential energy curves for the lowest singlet and triplet excited
ungerade states of the Sr$_2$  
molecule corresponding to the $^1{\rm S + {^3P}}$, $^1{\rm S + {^3D}}$, and
$^1{\rm S + {^1 D}}$ dissociation limits
have been obtained by a supermolecule method:
\be
V^{\rm ^{2S+1}|\Lambda|_u}(R)=
E_{\rm AB}^{\rm SM} -
E_{\rm A}^{\rm SM}-E_{\rm B}^{\rm SM},
\label{cccv}
\ee
where $E_{\rm AB}^{\rm SM}$ denotes
the energy of the dimer computed using the supermolecule method (SM), and $E_{\rm X}^{\rm SM}$,
X=A or B, is the energy of the atom X.
Here, the molecular electronic term is denoted by $^{2S+1}|\Lambda|_u$
where $S$ is the total electronic spin quantum number, and $\Lambda$ the projection of the
electronic orbital angular momentum on the molecular axis.
The excited states were calculated employing the linear response
theory (equation of motion) within the coupled-cluster singles, 
doubles, and linear triples (LRCC3) framework~\cite{Bartlett,Jorgensen:90,Helgaker:97}.
Note that in Refs. \cite{Moszynski:03,Moszynski:05} the full configuration
interaction correction (FCI) for the four-electron valence-valence
correlation was added on top of the linear response
result. However, the results for Ca$_2$
\cite{Moszynski:03,Moszynski:05} and  more recently  on Mg$_2$ show
that for the states of interest this correction is so small that it
can safely be neglected. We assume that this is also the 
case for Sr$_2$.

Transitions from the ground X$^1\Sigma_g^+$ state to the
$^1\Sigma_u^+$ and $^1\Pi_u$ states are electric dipole allowed.
The corresponding transition dipole moments are given by the
following expression 
\cite{Bunker:98}:
\be
\mu_i(n\leftarrow {\rm X})=\langle{\rm X}^1\Sigma_g^+|r_i|
(n)^1|\Lambda|_u\rangle,
\label{trandipel}
\ee
where $n$ numbers the consecutive nonrelativistic dissociation limits
of the $^1|\Lambda|_u$ states, and 
$r_i$, $i=x,y$ or $z$, denotes the $i$th component of the position vector.
Note that in Eq. (\ref{trandipel}), $i=x$ or $y$ corresponds to transitions
to $^1\Pi_u$ states, while $i=z$ corresponds to transitions to $^1\Sigma_u^+$
states.  In the present calculations the electric transition dipole moments
were computed as the first residue
of the coupled-cluster linear response function
restricted to single and double excitations (LRCCSD)
with two electric dipole operators~\cite{Jorgensen:90}. Note that in principle the more
advanced LRCC3 method could be used to calculate the transition moments. 
However, since the intensities in the spectra cannot be measured to a
high precision, we have employed 
a less accurate method limited to single and double excitations.
Comparison of the computed and measured atomic lifetimes will show
below that such an approximation is  sufficient for the purpose of the
present study.

The rovibrational energy levels of the electronically excited states 
of Sr$_2$ are expected to show some perturbations due to the
nonadiabatic coupling between 
the electronic states. Analysis of the potential energy curves, cf.
sec. \ref{sec3}, reveals angular couplings of the a$^3\Sigma_u^+$ and
b$^3\Sigma_u^+$ states with the c$^3\Pi_u$ state. 
Therefore, in this work we have computed the most
important angular coupling matrix elements defined by the expression:
\be
L({n\leftrightarrow n'})=\langle(n)^{2S+1}|\Lambda|_u|L_{\pm}|(n')^{2S+1}|\Lambda'|_u\rangle,
\label{ang}
\ee
with $L_{\pm}$ the ladder operator of the electronic angular momentum  and
$n\leftrightarrow n'$ denoting the coupling between
electronic states $n$ and $n'$ (here
$n$ will stand for the a$^3\Sigma_u^+$ or b$^3\Sigma_u^+$ states, 
$n'$ for the c$^3\Pi_u$ state).
Note that the electronic angular
momentum operator couples states with the projection of the
electronic orbital angular momentum on the molecular axis $\Lambda$ differing by one.
In the present calculations the angular coupling between the triplet states
was computed directly from the multireference configuration interaction
wave functions limited to single and double excitations (MRCI).

Strontium is a heavy atom, and the electronic states of the
Sr$_2$ molecule are expected to be  strongly mixed by the 
spin-orbit  interaction. Therefore, the spin-orbit coupling and its dependence on the 
internuclear distance $R$ must be taken into account in our analysis of the spectra
in the mixed singlet/triplet A$^1\Sigma_u^+$, c$^3\Pi_u$, and a$^3\Sigma_u^+$
manifold of electronic states. 
We have evaluated the spin-orbit coupling matrix elements
for the lowest dimer states that couple to the 0$^+_u$ and $1_u$ states of Sr$_2$
within the MRCI framework.
In our case the nonrelativistic states that are coupled through the spin-orbit
interaction to the $0_u^+$ symmetry are c$^3\Pi_u$, A$^1\Sigma_u^+$
and  ${\rm B}^1\Sigma_u^+$ \cite{Moszynski:06a,jmbrown:03}.
Other electronic states of Sr$_2$ can be omitted from the present analysis 
due to their very weak couplings with the A+c+a manifold and significant energetic gaps
as compared to these electronic states. Thus, the most important 
spin-orbit coupling matrix elements for states of the 0$^+_u$ symmetry
are given by 
\be
\begin{split}
& A(R) =\\&
 \langle {\rm c}^3\Pi_u(\Sigma=\pm1,\Lambda=\mp1)|
\widehat{H}_{\rm SO}|{\rm c}^3\Pi_u(\Sigma=\pm1,\Lambda=\mp1)\rangle\,,
\end{split}
\label{aso}  
\ee
\be
\xi_1(R) = \langle {\rm c}^3\Pi_u(\Sigma=\pm1,\Lambda=\mp1)|
\widehat{H}_{\rm SO}|{\rm A}^1\Sigma_u^+\rangle\,,
\label{xi1}
\ee
\be
\xi_2(R) =\langle {\rm c}^3\Pi_u(\Sigma=\pm1,\Lambda=\mp1)|
\widehat{H}_{\rm SO}|{\rm B}^1\Sigma_u^+\rangle\,,
\label{xi2}
\ee
where $\widehat{H}_{\rm SO}$ is the spin-orbit Hamiltonian in the
Breit-Pauli approximation \cite{Bethe:57} and 
$\Sigma$ denotes the projection of the electron spin angular
momentum on the molecular axis.
For the $1_u$ symmetry the most important couplings occur between the
c$^3\Pi_u$, ${\rm a}^3\Sigma_u^+$, ${\rm b}^3\Sigma_u^+$ and ${\rm B}^\prime {^1\Pi_u}$ states
\cite{Moszynski:06a,jmbrown:03}, and the corresponding spin-orbit matrix elements read:
\be
\varphi_1(R) = \langle {\rm c}^3\Pi_u(\Sigma=0,\Lambda=\pm1)|
\widehat{H}_{\rm SO}|{\rm a}^3\Sigma_u^+(\Sigma=\pm1,\Lambda=0)\rangle\,,
\label{phi1}
\ee
\be
\varphi_2(R) = \langle {\rm c}^3\Pi_u(\Sigma=0,\Lambda=\pm1)|
\widehat{H}_{\rm SO}|{\rm b}^3\Sigma_u^+(\Sigma=\pm1,\Lambda=0)\rangle\,,
\ee
\be
\varphi_3(R) = \langle {\rm c}^3\Pi_u(\Sigma=0,\Lambda=\pm1)|
\widehat{H}_{\rm SO}|{{\rm B}^\prime} ^1\Pi_u\rangle\,,
\ee
\be
\zeta_1(R) = \langle {\rm a}^3\Sigma_u^+(\Sigma=\pm1,\Lambda=0)|
\widehat{H}_{\rm SO}|{{\rm B}^\prime} ^1\Pi_u\rangle\,,
\ee
\be
\zeta_2(R) = \langle {\rm b}^3\Sigma_u^+(\Sigma=\pm1,\Lambda=0)|
\widehat{H}_{\rm SO}|{{\rm B}^\prime} ^1\Pi_u\rangle\,.
\label{zeta2}
\ee

With the spin-orbit coupling matrix elements at hand, we can build up the
matrices that will generate the potential energies  of the spin-orbit  states
that couple to  0$^+_u$ and 1$_u$ symmetry.
The matrices for the $0_u^+$ and $1_u$ states are  given by:
\begin{equation} 
{\Bbb V}^{0_u^+}=
\left( \begin{array}{ccc}
V^{{\rm c}^3\Pi_u}(R)-A(R) & \xi_1(R) & \xi_2(R) \\
\xi_1(R) & V^{{\rm A}^1\Sigma_u^+}(R) & 0 \\
\xi_2(R) & 0 & V^{{\rm B}^1\Sigma_u^+}(R) 
\end{array} \right)
\label{SO0}
\end{equation}
and 
\begin{equation} 
{\Bbb V}^{1_u}=
\left( \begin{array}{cccc}
V^{{\rm a}^3\Sigma_u^+} & 0 & \varphi_1(R) & \zeta_1(R) \\ 
0 & V^{{\rm b}^3\Sigma_u^+} & \varphi_2(R)  & \zeta_2(R)  \\ 
\varphi_1(R) & \varphi_2(R)  & V^{{\rm c}^3\Pi_u}(R) &  \varphi_3(R) \\
\zeta_1(R) & \zeta_2(R) & \varphi_3(R) & V^{{{\rm B}^\prime} ^1\Pi_u}(R) 
\end{array} \right),
\label{SO1}
\end{equation}
respectively.
\begin{table*}[ht!]
\caption{
Parameters of the analytical fits  of the a$^3\Sigma_u^+$,
b$^3\Sigma_u^+$, c$^3\Pi_u$, and A$^1\Sigma_u^+$ potentials (in atomic
units) for Sr$_2$. Numbers in parentheses denote the power of 10.
\label{tab1}
}
\vskip 5ex
\begin{tabular}{c @{\hspace{-1.5cm}}   d @{\hspace{-1.5cm}}   d @{\hspace{-1.5cm}}   d  @{\hspace{-1.5cm}}  d }
\hline\hline
        Parameter     & \multicolumn{1}{r}{\hspace{3cm}a$^3\Sigma_u^+$}  &  \multicolumn{1}{r}{b$^3\Sigma_u^+$}  &  \multicolumn{1}{r}{c$^3\Pi_u$}     & \multicolumn{1}{c}{\hspace{0.5cm}A$^1\Sigma_u^+$}    \\
          \hline
         $A_{0}$      &6.763580846(2)       &3.812370308(5)      &5.806031460(6)     &1.0713571607(3)      \\
         $A_{1}$      &-2.7286531831(2)    &-2.7785352110(5)   &-3.433599894(6)  &-3.1369577995(2)     \\
         $A_{2}$      &4.4648343035(1)      &7.599745922(4)      &7.677597973(5)     &4.259633532(1)        \\
         $A_{3}$      &-3.4042105677    &-9.26259308(3)     &-7.75564134(4)    &-2.828788402      \\
         $A_{4}$      &0.1048331384      &4.2607133921(2)     &3.0330564923(3)    &0.08207453705      \\
         $\alpha$     &0.03722198867     &0.7652092533     &2.102245892    &0.2810137665       \\
         $\beta$      &3.14301181557      &0.7516213273     &1.03238202    &1.195331858        \\
         $\gamma$     &0.07613828056         &0.148486204      &1.349678173(-3) &0.0362537402       \\
         $C_{12}$     &-5.318418476(9)     &1.1998965806(11)     &-1.06415514(10)   &7.278665842(11)       \\
         $C_{5}$      &                         &                        &                       &-8.649(2)               \\
         $C_{6}$      &4.488(3)                &6.750(3)               &3.951(3)              &2.72(3)                     \\
         $C_{8}$      &1.426(6)                &2.044(4)               &3.521(5)              &2.285(5)                 \\
         $C_{10}$     &2.321(8)                &1.010(6)               &3.296(7)              &9.223(7)                 \\
\hline\hline
\end{tabular}
\end{table*}
Diagonalization of these matrices yields the spin-orbit coupled potential
energy curves for the $0^+_u$ and $1_u$ states. Note that
all the potentials in the matrices (\ref{SO0}) and (\ref{SO1}) are
taken from the LRCC3 calculations. Only the
diagonal and nondiagonal spin-orbit coupling matrix elements were
obtained with the MRCI method.
Once the eigenvectors of these matrices are available, one can easily
get the electric dipole transition moments and the nonadiabatic coupling
matrix elements between the relativistic states.
It is worth noting that here,  unlike in the case of Ca$_2$, the
B$^1\Sigma_u^+$ and B$^\prime$$^1\Pi_u$ states are included in the
model. This is due to the fact that for Sr$_2$ the long-range
spin-orbit interactions of the c$^3\Pi_u$ state with the B and
B$^\prime$  states 
have some significance  since they are responsible for  the existence of very weakly bound states
located just below the $^1{\rm S_0 + {^3P_1}}$ threshold that were observed 
in the photoassociation experiment by Zelevinsky {\it et al.}~\cite{ZelevinskyPRL06}.
The long-range character of these couplings makes the photoassociation of the
ultracold strontium atoms possible, yielding a non-negligible resonant
$\delta C_3^{\rm res}R^{-3}$ interaction for the 0$_u^+$ and 1$_u$ potentials
at the $^1{\rm S_0 + {^3P_1}}$ asymptote, 
and a non-negligible relativistic transition
moment from the X$^1\Sigma_g^+$ ground state. By contrast, the
coupling with the ${{\rm A}^\prime}^1\Pi_u$ 
state which was reported in Ref. \cite{Tiemann:11} was neglected in the present calculations 
since it is asymptotically zero for the
$1_u$ states of interest and influences the $R^{-6}$ and higher
asymptotics of the $(1)1_u$ state,
as opposed to the $\delta C_3^{\rm res}R^{-3}$ asymptotics due to  
spin-orbit coupling with the ${{\rm B}^\prime}^1\Pi_u$ state,
cf. Ref.~\cite{JulienneEPL} for a simple atomic model and
Ref.~\cite{Skomorowski:un} for a rigorous explanation. 
Note finally that the spin-orbit coupling between the 
c$^3\Pi_u$, ${\rm a}^3\Sigma_u^+$, ${\rm b}^3\Sigma_u^+$ and ${\rm B}^\prime {^1\Pi_u}$ states
also leads to states
of $0_u^-$ and $2_u$ symmetry. In the absence of strong 
nonadiabatic effects these states are not optically active and we do
not discuss them here. 

In order to mimic the scalar relativistic effects, some electrons were described
by the ECP28MDF pseudopotential \cite{Stoll:06} from the Stuttgart library. 
Thus, in the present study the Sr$_2$ molecule was treated as a system of effectively 20 electrons.
In all calculations the  $[8s8p5d4f1g]$ basis set 
suggested in Ref. \cite{Stoll:06} was used, augmented
by a set of $[1s1p1d1f3g]$ diffuse functions. In the
calculations of the potentials this basis was supplemented by a
set of $spdfg$ bond functions \cite{Partridge:99}.
The full basis of the dimer was employed in the supermolecule
calculations and the Boys and Bernardi scheme was utilized
to correct for the basis-set superposition error \cite{Boys:70}.
{\em Ab initio} calculations were performed for a set of 22
interatomic distances ranging from $R$ = 5 to 50 bohr. In the calculations
of the potential energy curves and transition dipole moments the
{\sc dalton} code \cite{dalton20} was used. All MRCI calculations were done with the
{\sc molpro} code \cite{molpro06}. 
We would like to emphasize that almost all {\em ab initio}
results were obtained with the most advanced size-consistent methods of quantum
chemistry, LRCC3 and LRCCSD. Only the spin-orbit coupling matrix elements
and nonadiabatic matrix elements were obtained with the MRCI method which is
not size consistent. Fortunately, all of the couplings are important
in the region of the curve crossings or avoided crossings at short interatomic distances,
so the effect of the size-inconsistency of MRCI on our results should
not be dramatic.

\subsection{Analytical fits}
\label{sub2}
The computed points of the potential energy curves were fitted to the following 
analytical expression:
\begin{eqnarray}
\nonumber
V^{\rm ^{2S+1}|\Lambda|_u}(R) &=& e^{-\alpha R-\gamma R^2}\sum_{i=0}^4A_iR^i 
+ \frac{C_k^{\rm res}}{R^k}f_k(\beta,R)\\ &&-\sum_{n=3}^6\frac{C_{2n}}{R^{2n}}f_{2n}(\beta,R),
\label{fitex}
\end{eqnarray}
where $\{A_i\}_{i=0}^4$, $\alpha$,  $\beta$, $\gamma$, and $C_{12}$ were adjusted to the
computed points. The damping function $f_n(\beta,R)$ was employed in the form
proposed by Tang and Toennies \cite{Toennies}. 
The long-range coefficients $\{C_{2n}\}_{n=3}^5$ 
were not fitted, but fixed at the
{\em ab initio} values taken from Ref. \cite{Mitroy:2010}.
In our case the leading long-range coefficient $C_k^{\rm res}$ describes the
first-order resonant interaction \cite{Meath:68}
between the Sr($^1$S) and Sr($^1$D) atoms, and was also fixed at the
{\em ab initio} value \cite{Mitroy:2010}. For the triplet states the resonant
interaction term vanishes identically in the nonrelativistic approximation, 
so $C_k^{\rm res}$ is equal to zero. The parameters of the analytical fits of the
potentials are reported in Table \ref{tab1}.

The transition moments $\mu_0^{n\leftarrow {\rm X}}(R)$, spin-orbit
coupling matrix elements, $A(R)$, $\{\xi_i(R)\}_{i=1}^2$,
$\{\zeta_i(R)\}_{i=1,2}$, $\{\varphi(R)\}_{i=1}^3$, and  
angular coupling matrix elements, 
$L({\rm a\leftrightarrow c})$ and $L({\rm b\leftrightarrow c})$, were
fitted to the following generic expression:
\begin{eqnarray}
\nonumber
X(R) &=& X^\infty  
+ (A_0^X+A_1^X R+A_2^XR^2)e^{-\alpha_2^X R-\gamma^XR^2}\\
&&+B^Xe^{-\alpha_1^X R}   
+\sum_{n=n_0}^{6}\frac{X_{n}}{R^{n}}f_{n}(\beta^X,R),
\label{fitmom}
\end{eqnarray}
where $X$ stands for $\mu_0^{ n \leftarrow \rm{X}}(R)$, 
$A(R)$, $\{\xi_i(R)\}_{i=1}^2$, $\{\zeta_i(R)\}_{i=1}^2$, $\{\varphi(R)\}_{i=1}^3$, 
$L({\rm a\leftrightarrow c})$, and $L({\rm b\leftrightarrow c})$.
The leading
power in the inverse power expansion of Eq.~(\ref{fitmom}) depends on the 
asymptotic multipole expansion
of the wave functions in the polarization approximation
\cite{Avoird:77,Cwiok:92,Heijmen:96} and of the appropriate operator
and varies between 3 and 6 for the different quantities $X$.
The atomic values, $X^\infty$, were fixed as follows:
\be
\begin{split}
&A^\infty=\varphi_1^\infty=193.68\;{\rm cm}^{-1},\\
&\varphi_3^\infty=-\zeta_1^\infty=\xi_2^\infty/\sqrt{2}=-153.02\;{\rm cm}^{-1},
\end{split}
\ee
\be
\xi_1^\infty = \varphi_2^\infty = \zeta_2^\infty = 0,
\ee
\be
\mu_0^\infty({\rm A\leftarrow X}) = 0,
\; \; \; \; \;
\mu_0^\infty({\rm B\leftarrow X}) = \sqrt{2}\cdot3.07\;{\rm a.u.} ,
\label{Axiinf}
\ee
\be
L^\infty({\rm a\leftrightarrow c}) = \sqrt{2},
\; \; \; \; \;
L^\infty({\rm b\leftrightarrow c}) = 0,
\ee
and  the remaining parameters were adjusted to the
{\em ab initio} points. The value for $A^\infty$ was derived from
the experimental positions of the states in the $^3$P multiplet 
assuming pure $LS$ coupling, while  the values of $\xi_2^\infty$ and $\mu_0^\infty({\rm B\leftarrow X})$ are
based on the present {\em ab initio} atomic calculations. 
The parameters of the analytical fits of the
most important spin-orbit coupling matrix elements $A(R)$, $\xi_1(R)$, and
$\xi_2(R)$ are reported in Table \ref{tab1a}. All parameters for other fitted
quantities can be obtained from the authors upon request.

Note that fixing our fits at their proper
asymptotic values is crucial for a proper description of the rovibrational transitions
near the dissociation threshold. This is in a sharp contrast with some
potentials fitted to 
the experimental data that may not be sufficiently sensitive to the long-range tail
of the potential and spin-orbit couplings. 
{\small
\begin{table}[ht!]
\caption{
Parameters of the analytical fits of the spin-orbit coupling matrix elements (in atomic units) for Sr$_2$. Numbers in parentheses denote the power of 10.
\label{tab1a}
}
\vskip 5ex
\begin{tabular}{l @{\hspace{-2.0cm}} d @{\hspace{-1.5cm}} d @{\hspace{-1.4cm}}  d}
\hline\hline
  Parameter                &  A(R){\hspace{-1.0cm}}        &  \xi_1(R){\hspace{-1.0cm}}  &  \xi_2(R){\hspace{-1.0cm}}       \\
    \hline
         $X^\infty   $      &8.824756(-4)      &                                 &-9.860287(-4) \\
         $B^X        $      &                            &0.1281809               &-8.7008164(-3)  \\
         $\alpha_1^X $      &                            &0.255215595               &0.40876725       \\
         $A_0^X      $      &0.254210634           &-8.18616724(-2)         &2.96378955(-6)  \\
         $A_1^X      $      &-3.332647733(-2)    &9.17710750(-3)           &-3.39493117(-7) \\
         $A_2^X      $      &                            &-4.77060945(-4)         &1.0586325(-8)  \\
         $\alpha_2^X $      &0.7905806772          &                                 &-0.92238515      \\
         $\gamma^X   $      &                            &1.40267678(-2)          &2.8575636(-2)  \\
         $\beta^X    $      &1.3932739215          &6.8008294                &2.22429216     \\
         $X_3        $      &                            &                                 &2.64888569(-2)      \\
         $X_4        $      &                            &-1.684296847             &      \\
         $X_6        $      &-35.590169218        &                                 &      \\
\hline\hline
\end{tabular}
\end{table}
}

\subsection{Quantum-dynamical calculations}
\label{sub3}
In the present paper we consider the homonuclear bosonic $^{88}$Sr$_2$ molecule.
The rovibrational energy levels and wave functions for the ground X$^1\Sigma_g^+$ state were 
obtained by diagonalizing the Hamiltonian for the nuclear motion in the Born-Oppenheimer 
approximation with the variable step-size Fourier grid  representation
\cite{SlavaJCP99,WillnerJCP04,ShimshonCPL06}. For the ground state, an accurate 
potential  fitted to the experimental high-resolution Fourier transform  
spectra is available \cite{Tiemann:10} and is used  in our calculations.

Due to the spin-statistical weights for the bosonic $^{88}$Sr$_2$ molecule, we can limit
ourselves to odd values of the rotational quantum number $J$ and $e$ parity levels \cite{Bunker:98}. 
Rovibrational energy levels for the excited electronic states in the 
A$^1\Sigma_u^+$, c$^3\Pi_u$, and a$^3\Sigma_u^+$ manifold were obtained by diagonalizing the
following Hamiltonian:
\begin{widetext}
\be
\hspace{-1cm}\widehat{{\Bbb H}}=
\left( 
\begin{array}{ccccccc}
\widehat{H}^{{\rm c}^3\Pi_u}_{\rm diag}-A(R)&   \xi_1                 & \xi_2            & -C_{JL}    &  -C_{JL}   & C_{JS}            & 0                       \\
\xi_1                      & \widehat{H}^{{\rm A}^1\Sigma_u^+}_{\rm diag}& 0                   & 0                & 0                &0                 & \sqrt{2}C_{JL}           \\
\xi_2                      & 0                 & \widehat{H}^{{\rm B}^1\Sigma_u^+}_{\rm diag}& 0                & 0                &0                 & \sqrt{2}C_{JL}           \\
-C_{JL}              & 0                 & 0                   &\widehat{H}^{{\rm a}^3\Sigma_u^+}_{\rm diag} & 0                & \varphi_1+C_{LS}& \zeta_1              \\ 
-C_{JL}              & 0                 & 0                   &0                 & \widehat{H}^{{\rm b}^3\Sigma_u^+}_{\rm diag}& \varphi_2+C_{LS}& \zeta_2              \\ 
C_{JS}              & 0                 & 0                   &\varphi_1+C_{LS}& \varphi_2+C_{LS}& \widehat{H}^{{\rm c}^3\Pi_u}_{\rm diag}&  \varphi_3           \\
0                          & \sqrt{2}C_{JL}     & \sqrt{2}C_{JL}       &\zeta_1        & \zeta_2       & \varphi_3     & \widehat{H}^{{{\rm B}^\prime} ^1\Pi_u}_{\rm diag} 
\end{array} \right),
\label{HAMp1}
\ee
where the diagonal term is given by:
\begin{equation}
\widehat{H}_{\rm diag}^{(n)^{(2S+1)}|\Lambda|_u}  \equiv
\frac{1}{2\mu}\widehat{p}_R^2 + V_{n}^{^{(2S+1)}|\Lambda|_u}(R) +
\frac{J(J+1) + S(S+1) +L(L+1) - \Omega^2 -\Sigma^2-\Lambda^2}
{2\mu R^2}
\label{Vdiag}
\end{equation}
\end{widetext}
with $\widehat{p}_R$ denoting the radial momentum operator, $\mu$ the
reduced mass of the dimer, and $\Omega$ is the projection
of the total electronic angular momentum on the molecular axis,
so $\Omega=\Lambda+\Sigma$.  
The first three rows of the $\widehat{{\Bbb H}}$  matrix correspond
to the states with $|\Omega| = 0$ while the last four rows to the states
with $|\Omega| = 1$.
The angular (Coriolis-type) couplings are defined as:
\be
C_{JL}(R)=-\frac{\left[J(J+1)\right]^{1/2}L(R)}
{2\mu R^2},
\label{CJL0}
\ee
\be
C_{JS}(R)=-\frac{\left[2J(J+1)\right]^{1/2}}
{2\mu R^2},
\label{CJS0}
\ee
\be
C_{LS}(R)=2^{1/2}\frac{L(R)}
{2\mu R^2},
\label{CLS}
\ee
where 
$L(R)$ stands for $L({n\leftrightarrow n'})$ defined by Eq. (\ref{ang}) with $n$ and
$n'$ properly chosen, the quantum number $L$
appearing in Eq. (\ref{Vdiag}) is the electronic angular quantum number of 
the excited state atom,
and all other symbols appearing in Eq. (\ref{HAMp1}) are defined in  
Eqs. (\ref{aso})--(\ref{xi2}) and (\ref{phi1})--(\ref{zeta2}). We
refer the reader to Ref. \cite{Moszynski:06a} for a rigorous 
justification of the above expressions. Note that the term
$\frac{L(L+1)}{2\mu R^2}$ is not 
rigorously correct, since it results from the so-called adiabatic (diagonal) correction
for the nuclear motion and the above mentioned value is true only in the separated atoms
limit. At present, there is no {\em ab initio} electronic structure code 
that could provide us with (even approximate) values of the angular part of the adiabatic
correction, so we keep it at its asymptotic atomic value. This approximation
should work very well for the rovibrational levels near the dissociation threshold.
If the nonadiabatic angular coupling matrix
elements are small, we can set the Coriolis coupling constants $C_{JL}$, $C_{JS}$, and
$C_{LS}$ equal to zero, and  the matrix $\widehat{{\Bbb H}}$ 
becomes block diagonal with two blocks corresponding
separately to the $0_u^+$ and $1_u$ levels:
\be
\widehat{{\Bbb H}}^{0_u^+}=
\left( 
\begin{array}{ccc}
\widehat{H}^{{\rm c}^3\Pi_u}_{\rm diag}-A(R)&   \xi_1                 & \xi_2                     \\
\xi_1                      & \widehat{H}^{{\rm A}^1\Sigma_u^+}_{\rm diag}& 0                     \\
\xi_2                      & 0                 & \widehat{H}^{{\rm B}^1\Sigma_u^+}_{\rm diag}    \\
\end{array} \right),
\label{rovibmat0}
\ee
\be
\widehat{{\Bbb H}}^{1_u}=
\left( 
\begin{array}{cccc}
\widehat{H}^{{\rm a}^3\Sigma_u^+}_{\rm diag} & 0                & \varphi_1& \zeta_1              \\ 
0                 & \widehat{H}^{{\rm b}^3\Sigma_u^+}_{\rm diag}& \varphi_2& \zeta_2              \\ 
\varphi_1& \varphi_2& \widehat{H}^{{\rm c}^3\Pi_u}_{\rm diag}&  \varphi_3           \\
\zeta_1        & \zeta_2       & \varphi_3     & \widehat{H}^{{{\rm B}^\prime} ^1\Pi_u}_{\rm diag} 
\end{array} \right).
\label{rovibmat1}
\ee

The line strength in the spectra, $S(v'J'\leftarrow v''J'')$, from the rovibrational
level $|v'',J'' \rangle$ of the ground electronic state to the rovibrational level $|v',J'\rangle$ 
of the A$^1\Sigma_u^+$, c$^3\Pi_u$, and a$^3\Sigma_u^+$ manifold is given by:
\be
\begin{split}
&S(v'J'\leftarrow v''J'')=(2J'+1)H_{J'}\\
&\times\Big|\sum_{\Omega'=0,\pm 1}
\sum_{\sigma=-1}^1
\sum_{n'}
\langle\chi_{1J''0}(v'')|
r_\sigma(n'\leftarrow {\rm X})|
\chi_{n'J'\Omega'}(v')\rangle\Big|^2,
\label{SfinBO}
\end{split}
\ee
where $H_{J'}$ is the so-called H\"onl-London factor, 
\be
H_{J'}=
\begin{cases}\frac{J'+1}{2J'+1} & \text{for} \quad J'=J''-1, \\
 \frac{1}{2J'+1}  & \text{for} \quad J'=J'', \\
 \frac{J'}{2J'+1} & \text{for} \quad J'=J''+1,
\end{cases}
\label{HL}
\ee
and $\chi_{1J''0}(v'')$ is the rovibrational wave function of the ground state,
while $\chi_{n'J'\Omega'}(v')$ is an eigenfunction of the 
Hamiltonian, either $\widehat{{\Bbb H}}$ given by Eq. (\ref{HAMp1}),
or $\widehat{{\Bbb H}}^{0_u^+}$ and  $\widehat{{\Bbb H}}^{1_u}$ given
by Eqs. (\ref{rovibmat0}) and 
(\ref{rovibmat1}), if the helicity decoupling approximation
(neglecting of the angular Coriolis couplings) is employed.

We also study the positions and lifetimes of Feshbach resonances
appearing just above the ${\rm ^1S_0\;+\;^3P_1}$ dissociation
threshold. The Feshbach resonances of interest are quasi-bound
rovibrational levels of the (2)$0_u^+$  
electronic state lying above the dissociation limit of the (1)$0_u^+$
state, i.e., in the continuum of this state.
Formally, the Feshbach resonances can be characterized by complex energies of the
form $E_r-(i/2)\Gamma$, where $E_r$ denotes the position of the
resonance state and 
$\Gamma$ its width which is directly related to the lifetime $\tau$ by
$\tau=\hbar/\Gamma$. 
We have determined these complex energies by diagonalizing
the Hamiltonian for the $0_u^+$ states, Eq. (\ref{rovibmat0}), with an imaginary
absorbing potential $V_{\rm CAP}$ added to the diagonal Hamiltonian
terms (\ref{Vdiag}) \cite{Meyer:93,Manolopoulos:02,Manolopoulos:04}:
\be
V_{\rm CAP}(R) =
\begin{cases}
0&\hspace{-0.3cm}, R \le R_{\rm a}, \\
\frac{4E_{\rm min}}{C^2}\left[\frac{1}{(1-x)^2}+\frac{1}{(1+x)^2}-2 \right]&\hspace{-0.3cm}, R_{\rm a} \le R < R_{\rm max},
\end{cases}
\label{CAP}
\ee
where $C=2.62206$ and $x=(R-R_{\rm a})/(R_{\rm max}-R_{\rm a})$. The
parameters $E_{\rm min}$, $R_{\rm a}$, and  $R_{\rm max}$ were adjusted to 
obtain stable results with respect to small variations of these
parameters. 
Approximate positions of the resonances were first determined by using
the stabilization method \cite{Taylor:66} with respect to 
the size of the grid. With these positions at hand, a set 
of parameters leading to stable complex eigenvalues corresponding to the positions 
and widths of the resonances was easily found, yielding 
$R_{\rm a}=30$ bohr, $R_{\rm max}=200$ bohr, and $E_{\rm min}=30$ cm$^{-1}$.  
The knowledge of the approximate positions was 
particularly useful to determine the value of $E_{\rm min}$.

\section{Numerical results and discussion}
\label{sec3}
\subsection{{\em Ab initio} electronic structure data}
\label{sub4}
\begin{table}[ht!]
\caption{
Excitation energies (in cm$^{-1}$) for the low-lying energy levels of strontium atom.
\label{tab7}}
\vskip 5ex
\begin{tabular}{c @{\hspace{0.7cm}}  c  @{\hspace{0.7cm}} c  @{\hspace{0.7cm}}  c }
\hline\hline
Excited state &   Present   &  Reference \cite{Porsev:08} & Experiment \cite{nist}      \\
    \hline
$^3$P$_0$            &    14187.3   & 14241 &    14317.5   \\
$^3$P$_1$            &    14372.7   & 14448 &    14504.4   \\
$^3$P$_2$            &    14762.6   & 14825 &    14898.6   \\
$^3$D$_1$            &    18582.9   & 18076 &    18159.1   \\
$^3$D$_2$            &    18637.2   & 18141 &    18218.8   \\
$^3$D$_3$            &    18726.0   & 18254 &    18319.3   \\
$^1$D$_2$            &    20650.3   & 19968 &    20149.7   \\
$^1$P$_1$            &    21764.3   & 21469 &    21698.5   \\
\hline\hline
\end{tabular}
\end{table}
Before discussing the potential energy curves, we first discuss the atomic
excitation energies obtained from the LRCC3 calculations and the atomic lifetimes. 
In Table \ref{tab7} we present 
calculated excitation energies in comparison with fully relativistic
atomic calculations of Porsev and 
collaborators \cite{Porsev:08} and experimental data. 
Our predicted position of the nonrelativistic $^3$P state is 14570.8  cm$^{-1}$,
to be compared with the experimental value of 14704.9 cm$^{-1}$ \cite{nist} deduced 
from the positions of the states in the $^3$P multiplet and the Land\'e rule.
For the $^1$P$_1$ state we obtain 21764.3 cm$^{-1}$, again in very good  
agreement with the experimental value of 21698.5 cm$^{-1}$ \cite{nist}. 
For the nonrelativistic $^3$D state we obtain 
18668.8 cm$^{-1}$, to be compared with 18255.2 cm$^{-1}$ \cite{nist}
deduced from the positions of the states in the $^3$D multiplet and the Land\'e rule.
Finally, our term energy for the $^1$D$_2$ state is 20650.3  cm$^{-1}$, again in a
satisfactory agreement with experiment, 20149.7 cm$^{-1}$ \cite{nist}.
The accuracy of the atomic spin-orbit couplings can be judged by comparing  the computed
and observed splittings of the energy levels in the $^3$P and $^3$D multiplets.
For the $^3$P multiplet, theoretical splittings between the $^3$P$_2$ and $^3$P$_1$, $^3$P$_2$ and $^3$P$_0$, 
and $^3$P$_1$ and $^3$P$_0$ levels amount to 389.9 cm$^{-1}$, 575.3 cm$^{-1}$, and 185.4 cm$^{-1}$, respectively,
to be compared with the experimental numbers, 394.2 cm$^{-1}$, 581.0 cm$^{-1}$,
and 186.8 cm$^{-1}$, respectively.
A somewhat less  good  agreement is observed for the $^3$D multiplet. 
The theoretical fine splittings for the $^3$D$_3$--$^3$D$_2$, $^3$D$_3$--$^3$D$_1$, and $^3$D$_2$--$^3$D$_1$
states  read  88.9 cm$^{-1}$, 143.1 cm$^{-1}$, and 54.2 cm$^{-1}$, respectively,
while the experimental numbers are  100.5 cm$^{-1}$, 160.2 cm$^{-1}$,
and 59.7 cm$^{-1}$, respectively. 
Finally, we also note that the lifetimes of the $^3$P$_1$, multiplet $^3$D, 
$^1$D$_2$, and $^1$P$_1$ states of Sr are accurately reproduced.
Our calculated lifetimes together with the most recent experimental
and other theoretical results are listed in Table \ref{tab6}. For the $^1$P$_1$ 
state we obtained 5.09 ns to be compared with the experimental value of 5.22(3) ns 
\cite{Nagel:05}. For $^3$P$_1$ the theoretical and experimental numbers
are 21.4$\,\mu$s  and 21.5(2)$\,\mu$s \cite{ZelevinskyPRL06}, respectively. 
For the D states we observe a slightly worse  agreement. The theoretical
lifetime of the $^1$D$_2$ state is  
0.23 ms, to be compared with the experimental value of 0.30 ms  \cite{Courtillot:05}.
The same numbers for the average multiplet $^3$D  are 2.72$\,\mu$s and
2.5(2) $\mu$s \cite{Redondo:04}. 
Such a good agreement between theory and experiment for the atoms gives us
confidence that the molecular results will be of similar accuracy,
i.e., 
at worst a few percent off from the exact results.
\begin{table}[ht!]
\caption{
Comparison of the present and most recent theoretical and experimental values of the lifetimes of
low-lying excited states of strontium atom.
\label{tab6}}
\vskip 5ex
\begin{tabular}{c @{\hspace{0.5cm}}  d  @{\hspace{0.5cm}} l}
\hline\hline
\clc{Excited state} &   \clc{ lifetime        }   &  \clc{Reference}       \\
    \hline
$^1$P$_1$            & 5.09 \;{\rm ns}         &        Present                                \\
                     & 5.38  \;{\rm ns}         &             Theory,   Ref. \cite{Porsev:08}  \\
                     & 5.35 \;{\rm ns}          &             Theory,   Ref. \cite{Mitroy:2010}   \\
                     & 5.22(3)  \;{\rm ns}      &             Experiment,   Ref.  \cite{Nagel:05}       \\
                     & 5.263(4) \;{\rm ns}      &             Experiment,   Ref.  \cite{Yasuda:06}      \\
$^3$P$_1$            & 21.40 \;\mu{\rm s}    &        Present                             \\
                     & 24.4  \;\mu{\rm s}  &            Theory,    Ref.  \cite{Greene:04}      \\
                     & 19.0  \;\mu{\rm s}  &            Theory,    Ref.  \cite{Porsev:01}      \\
                     & 21.5(2)\;\mu{\rm s} &            Experiment,    Ref.  \cite{ZelevinskyPRL06}    \\
$^1$D$_2$            & 0.23 \;{\rm ms}         &        Present                            \\
                     & 0.412(10) \;{\rm ms}    &          Experiment,      Ref. \cite{Husain:88}  \\
                     & 0.30  \;{\rm ms}        &          Experiment,      Ref. \cite{Courtillot:05}  \\
$\overline{^3\textrm{D}}$ & 2.72  \;\mu{\rm s}    &   Present                                \\
                          & 2.4   \;\mu{\rm s}    &        Theory,   Ref.        \cite{Porsev:08}     \\
                          & 2.5(2)\;\mu{\rm s}    &        Experiment,   Ref.        \cite{Redondo:04}   \\
\hline\hline
\end{tabular}
\end{table}

One of the important issues in {\em ab initio} electronic structure calculations
is the quality of the basis set and of the wave functions. 
To further judge the quality of the basis set used in our calculations
we have computed the leading $C_6$ van der Waals coefficient for the ground
X$^1\Sigma_g^+$ state. This coefficient was obtained by using  the
explicitly connected representation of the expectation value
and polarization propagator within the coupled cluster method
\cite{Jeziorski:93,Moszynski:05a}, 
and the best approximation XCCSD4 proposed by Korona and collaborators
\cite{Korona:06}. Our {\em ab initio} result is 3142 a.u. which
compares very favorably with the value fitted to high-resolution
Fourier transform spectra, 3168(10) a.u. \cite{Tiemann:10}.  
The agreement between theory and experiment is better than
for most of the other {\em ab initio} calculations
\cite{Stanton:94,Lima:05,Mitroy:2003,Derevianko:2006,Mitroy:2010}.  
Comparison between theory and experiment for the 
well depth  of the ground state X$^1\Sigma_g^+$ potential is somewhat  less satisfactory. Our
theoretical value is  1124.0 cm$^{-1}$, to be compared
with the experimental result of 1081.64(2) cm$^{-1}$ \cite{Tiemann:10},
i.e., 3.8\% too large. 
However, in the case of the ground state interaction,
the FCI correction for the valence-valence correlation turned out to be
important. Due to computational limitations, we could
obtain it only in the Sadlej pVTZ basis \cite{Sadlej} which is comparatively small and does not allow for a better accuracy.
\begin{figure}[t!]
\begin{center}
\includegraphics[angle=-90,width=0.9\linewidth]{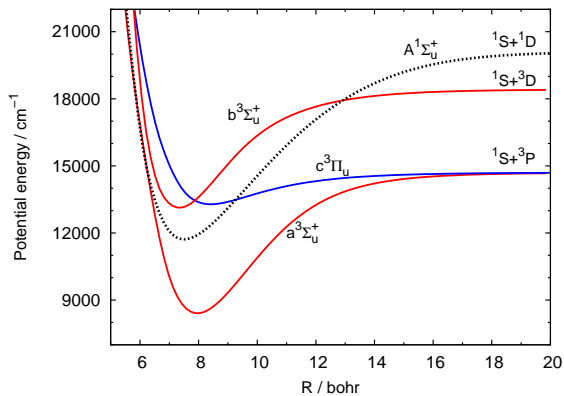}
\end{center}
\caption{
{\em Ab initio} potential energy curves for the A$^1\Sigma_u^+$, c$^3\Pi_u$,
a$^3\Sigma_u^+$, and b$^3\Sigma_u^+$ states of the strontium dimer.
}
\label{fig1}
\end{figure}
\begin{table}[b!]
\caption{
Spectroscopic characteristics of the non-relativistic electronic states
of Sr$_2$ dimer. \label{tab2}}
\vskip 5ex
\begin{tabular}{lcccc}
\hline\hline
  State \hspace{0cm} &  \hspace{0.3cm} $D_e$/cm$^{-1}$ \hspace{0.3cm} &  \hspace{0.3cm} $R_e$/bohr \hspace{0.3cm} & Ref. & Dissociation\\
    \hline
    a$^3\Sigma_u^+$   &   6298           &   7.95         &  Present & $^1$S+$^3$P  \\
                      &   6895           &   7.74         &  \cite{Czuchaj:03}  \\
                      &   6683           &   7.79         &  \cite{Frecon:96}  \\
    b$^3\Sigma_u^+$   &   5293           &   7.33         &  Present & $^1$S+$^3$D \\
                      &   4698           &   7.25         &  \cite{Czuchaj:03}  \\
                      &   5763           &   7.08         &  \cite{Frecon:96}  \\
    c$^3\Pi_u$        &   1422           &   8.41         &  Present & $^1$S+$^3$P  \\
                      &   1892           &   8.02         &  \cite{Czuchaj:03}  \\
                      &   1785           &   8.16         &  \cite{Frecon:96}  \\
    A$^1\Sigma_u^+$   &   8433           &   7.54         &  Present & $^1$S+$^1$D  \\
                      &   5440           &   7.12         &  \cite{Czuchaj:03}  \\
                      &   9066           &   7.28         &  \cite{Frecon:96}   \\
\hline\hline
\end{tabular}
\end{table}

The nonrelativistic potential energy curves relevant for the spectroscopy in the 
A+c+a manifold are plotted in Fig.~\ref{fig1}, while the spectroscopic characteristics 
of these states are reported in Table \ref{tab2}.
The separated atoms energy for each state was set equal to the experimental
value. Due to the computational limitations in the present work we did not consider 
the B$^1\Sigma_u^+$, A$^\prime$$^1\Pi_u$ and B$^\prime$$^1\Pi_u$ states.
Fortunately enough, the calculations of these potentials were not crucial for our
study. Indeed, the potential energy curve for the B$^1\Sigma_u^+$ state, 
with the correct $C_3^{\rm res}R^{-3}$ asymptotics,
could be fitted to the experimental data, and is available in the
literature \cite{Tiemann:11}. According to Ref. \cite{Tiemann:11}, the A$^\prime$$^1\Pi_u$
state is isolated, and its rovibrational levels are only weakly perturbed. Thus, it
can safely be omitted from our model for the rovibrational levels of the $1_u$ states
near the ${\rm ^1S_0 +{^3P_1}}$ threshold. However, would this state become interesting
from an experimental point of view, an accurate potential energy curve
fitted to the observed spectroscopic transitions is available in
Ref. \cite{Tiemann:11}, while the important spin-orbit coupling matrix elements can
be obtained from the present authors upon request.
The second state dissociating into the ${\rm ^1S +{^1P}}$ atoms, the B$^\prime$$^1\Pi_u$ state,
was not observed experimentally. In our work the potential energy curve for this
state was approximated with its long-range form $C_3^{\rm res}R^{-3}$,
with the proper long-range coefficient adapted from Ref. \cite{Mitroy:2010}.
While this is an approximation, it is not a crucial one, since the spin-orbit coupling of 
this state with the c$^3\Pi_u$ state is important only at large interatomic distances,
where it becomes constant, and affects 
the rovibrational dynamics of the Sr$_2$ molecule only near the intercombination
line $^1{\rm S_0 + {^3P_1}}$.
\begin{figure}[t!]
\begin{center}
 \includegraphics[angle=0,width=0.9\linewidth]{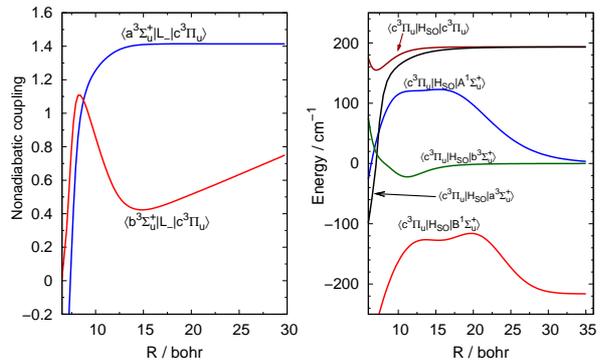}
\end{center}
\caption{
Nonadiabatic matrix elements (left-hand panel) and spin-orbit coupling matrix elements
(right-hand panel) as functions of the interatomic distance $R$.
}
\label{fig2}
\end{figure}

Let us compare our results with other available {\em ab initio} data 
from nonrelativistic calculations \cite{Czuchaj:03}.
The spectroscopic constants are listed in Table~\ref{tab2} and
compared to the results of Ref. \cite{Czuchaj:03}. Inspection of Table
\ref{tab2} shows that  the agreement with the data of Czuchaj {\it et
  al.} \cite{Czuchaj:03} is not satisfactory. For most of the states
the computed electronic binding energies agree  
within a few hundred cm$^{-1}$ at best, while the differences in the positions 
of the minima are 0.4  bohr at worst. 
The most striking difference between the present results and the data of Ref. 
\cite{Czuchaj:03} is the spectroscopically active A$^1\Sigma_u^+$ state.
Here, the difference in the position of the minimum is 0.42 bohr,
and the difference in $D_e$ is as much as 4000 cm$^{-1}$.
Surprisingly, the agreement of the present results with the older 1996
calculations by Aubert-Fr\'econ and collaborators \cite{Frecon:96} is
comparatively 
good. Except for the c$^3\Pi_u$ state, the well depths agree within 6\%
to 8\% and the well positions within 0.3 bohr at worst. For the
c$^3\Pi_u$ state we note a serious disagreement to all previous results. Our potential is
considerably shallower and the minimum is shifted to larger distances.
However, as will be shown in sec. \ref{sub5} the present picture of the interatomic
interactions in the A+c+a manifold reproduces all features of the
available experimental data.

The nonadiabatic  and spin-orbit coupling matrix elements 
as functions of the interatomic distance $R$ are reported in Fig. \ref{fig2}.
Note the maximum in the angular coupling between the c$^3\Pi_u$  and the
b$^3\Sigma_u^+$ states. The position of this maximum corresponds to the
crossing of the potential energy curves of these states. Also worth noting
is the broad maximum of the spin-orbit coupling between the A$^1\Sigma_u^+$
and c$^3\Pi_u$ states. This maximum extends to the region where the
potential energy curves cross, and is responsible for the strong mixing
of the singlet and triplet rovibrational energy levels. At large distances
this particular coupling tends to zero, but the coupling between the 
B$^1\Sigma_u^+$ and the c$^3\Pi_u$ states becomes important and is
responsible for the nonvanishing relativistic dipole moment
between the ground state and the triplet levels in the A+a+c manifold.
\begin{figure}[t!]
\begin{center}
\includegraphics[angle=-90,width=0.9\linewidth]{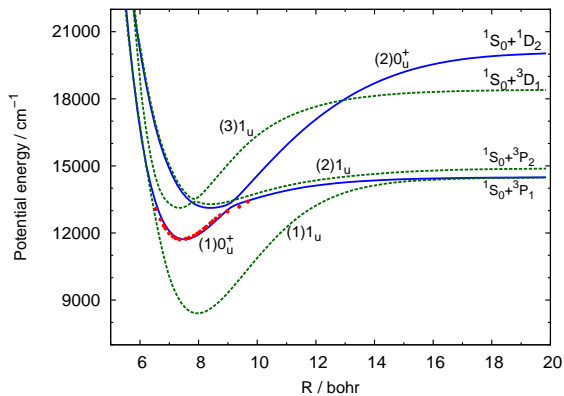}
\end{center}
\caption{
Relativistic potential energy curves for the lowest $0_u^+$ and $1_u$ states of the strontium dimer.
The dots represent the effective empirical potential fitted to the high-resolution Fourier
transform spectroscopic data of Ref. \cite{Tiemann:11}.
}
\label{fig3}
\end{figure}
\begin{figure}[b!]
\begin{center}
 \includegraphics[angle=0,width=0.9\linewidth]{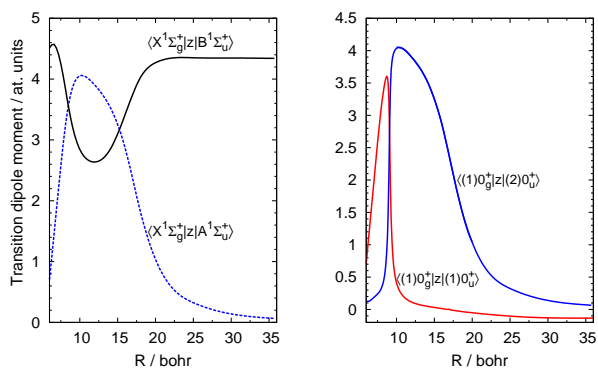}
\end{center}
\caption{
Transition dipole moments between the ground electronic state and the excited states of
Sr$_2$ in the non-relativistic basis (left-hand panel) and the relativistic basis (right-hand panel)
as functions of the interatomic distance $R$.
}
\label{fig4}
\end{figure}

The relativistic $0_u^+$ and $1_u$ potentials as functions of the
interatomic distance $R$ are depicted  in Fig. \ref{fig3}. Their
spectroscopic parameters are reported in Table \ref{tab3}. First,
we note an excellent agreement between the present {\em ab initio}
potential for the (1)$0_u^+$ state and the effective empirical 
potential fitted to the high-resolution Fourier
transform spectroscopic data \cite{Tiemann:11}. The well depths,
2782 cm$^{-1}$ on the theory side and 2790 cm$^{-1}$ from the fit
to the experimental data, agree within 8 cm$^{-1}$, while the well 
positions agree to within 0.05 bohr. It is gratifying to observe that the
present {\em ab initio} calculations also reproduce satisfactorily the
position and energy of the avoided crossing. Theory predicts
the avoided crossing between the (1) and (2)$0_u^+$ potentials
at $R$ = 9.1 bohr and $V$ = --1357 cm$^{-1}$, while the experimental
numbers are $R$ = 9.1 bohr and $V$ = --1470 cm$^{-1}$.
We note a substantial disagreement
between the present {\em ab initio} results and those reported
by Kotochigova \cite{Koto:08}. The difference in the well
depth is as large as 946 cm$^{-1}$. This means that 
Ref. \cite{Koto:08} does not predict any interaction between
the (1) and (2)$0_u^+$ states.  Thus, our result fully confirms 
the disagreement already noticed by Tiemann and collaborators \cite{Tiemann:11}.
For other states the agreement between the two calculations is quite erratic. 
For instance, the well depths for the second  state of $0_u^+$ symmetry and the
first state of $1_u$ symmetry differ substantially, while the
results for the $(2)1_u$ and $(3)1_u$ states are very close in
energy, but the positions of the wells are shifted by as much
as 0.5 to 0.9 bohr. 
\begin{table}[t!]
\caption{
Spectroscopic characteristics of the relativistic electronic states
of Sr$_2$ dimer. \label{tab3}}
\vskip 5ex
\begin{tabular}{lcccc}
\hline\hline
  State \hspace{0cm} &  \hspace{0.3cm} $D_e$/cm$^{-1}$ \hspace{0.3cm} &  \hspace{0.3cm} $R_e$/bohr \hspace{0.3cm} & Ref. & Dissociation\\
    \hline
    (1)0$_u^+$        &   2782           &   7.51         &  Present & $^1$S$_0$+$^3$P$_1$  \\
                      &   2790           &   7.46         &  \cite{Tiemann:11}  \\
                      &   1837           &   8.2          &  \cite{Koto:08}  \\
    (2)0$_u^+$        &   7039           &   8.39         &  Present & $^1$S$_0$+$^1$D$_2$ \\
                      &   5292           &   7.2          &  \cite{Koto:08}  \\
    (1)1$_u$          &   6097           &   7.95         &  Present & $^1$S$_0$+$^3$P$_1$  \\
                      &   6921           &   7.8          &  \cite{Koto:08}  \\
    (2)1$_u$          &   1942           &   7.34         &  Present & $^1$S$_0$+$^3$P$_2$  \\
                      &   1907           &   8.2          &  \cite{Koto:08}   \\
    (3)1$_u$          &   4863           &   7.97         &  Present & $^1$S$_0$+$^3$D$_1$  \\
                      &   4849           &   7.4          &  \cite{Koto:08}   \\
\hline\hline
\end{tabular}
\end{table}

Fig. \ref{fig4} shows the nonrelativistic electric transition
dipole moments from the ground state to the $^1\Sigma_u^+$ states and the relativistic
transition dipole moments to the $0_u^+$ states. Inspection
of Fig. \ref{fig4} reveals a broad maximum in the transition moment to the
A$^1\Sigma_u^+$ state at distances
around the minimum of the ground X$^1\Sigma_g^+$ state, and
decay to zero at large distances. The transition moment to the B$^1\Sigma_u^+$ state 
exhibits a broad minimum in the very same region and tends to the
atomic value at large distances. The two relativistic curves
reported in the right-hand panel of Fig. \ref{fig4}, when superimposed,
reproduce the transition moment to the A$^1\Sigma_u^+$ state.
This is not surprising, since at small distances the relativistic
transition moment to the (1)$0_u^+$ is dominated by the singlet
component. Around the crossing between the A$^1\Sigma_u^+$ and
the c$^3\Pi_u$ it drops off drastically due to the predominantly
triplet character of the electronic wave function. Note that at
large distances it does not decay to zero, but rather to a small
but constant atomic value reflecting the finite lifetime of the
atomic $^3$P$_1$ state. The opposite picture holds  for the
transition moment to the (2)$0_u^+$ state. At small distances
this state is dominated by the triplet component, and the
transition moment is very small. Starting from the curve
crossing, the (2)$0_u^+$ state is predominantly singlet, and
the transition dipole moment becomes  very large.

\subsection{Energy levels and rovibrational spectra of Sr$_2$ in
the A$^1\Sigma_u^+$, c$^3\Pi_u$, and a$^3\Sigma_u^+$ manifold}
\label{sub5}
Before comparing our results with the existing 
experimental data, let 
us discuss the validity of the decoupling between the $0_u^+$ and 
$1_u$ states. Similarly as for the Ca$_2$ \cite{Moszynski:06a} and SrYb \cite{Tomza:11} 
molecules, the nonadiabatic
angular coupling is completely negligible also for the strontium
dimer, except for a few most weakly bound levels. 
We have calculated the  bound states for all possible values
of the rotational quantum number $J$ by diagonalization of the
full Hamiltonian matrix (\ref{HAMp1}) with the variable step
Fourier grid representation, and submatrices corresponding
to the $0_u^+$ and $1_u$ blocks, Eqs. (\ref{rovibmat0})
and (\ref{rovibmat1}). We found that the eigenvalues of the decoupled 
matrices differ from the eigenvalues of the full matrix by 
less than $10^{-3}$ cm$^{-1}$, justifying the decoupled
representation. Actually, experimental data for the rovibrational energy
levels of the (1)$1_u$ state are scarce and limited to three
levels closest to the ${\rm ^1S_0 + {^3P_1}}$ dissociation limit,
so we will  focus our discussion on the $0_u^+$ levels.
\begin{figure}[b!]
\begin{center}
 \includegraphics[angle=0,width=1.0\linewidth]{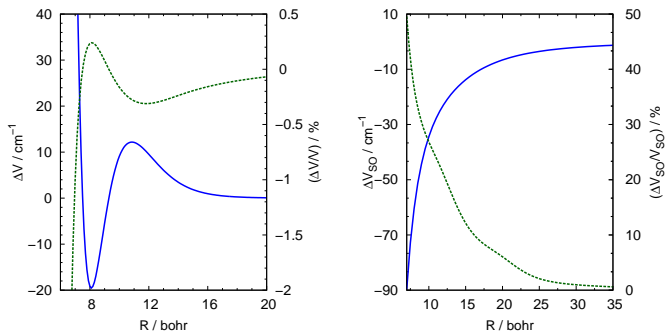}
\end{center}
\caption{
Left-hand panel:
Comparison of the original A$^1\Sigma_u^+$ potential with the fitted one
to the experimental data \cite{Tiemann:11} for $J'=1$. The solid blue line shows the absolute difference,
$\Delta V = V_{\rm fitted}$-$V_{ab\;\; initio}$, while the green dashed line shows the relative difference,
$\frac{\Delta V}{V}=\frac{V_{\rm fitted}-V_{ab\;\; initio}}{V_{ab\;\; initio}}\cdot 100\%$.
Right-hand panel: the same for the spin-orbit coupling term
$\xi_2=\langle {\rm c}^3\Pi_u |H_{\rm SO}|{\rm B}^1\Sigma_u^+ \rangle$.
}
\label{fig5}
\end{figure}

We have generated all energy levels for the $0_u^+$ and $1_u$
states up to and including $J'=$219. Comparison of the computed
energy levels for the $0_u^+$ state with the data derived
from the high-resolution Fourier transform spectroscopy show
a root-mean-square-deviation (RMSD) between theory and experiment
of 10.5 cm$^{-1}$. While such an agreement between theory and
experiment is very good for a system with 76 electrons, we
decided to adjust the {\em ab initio} data to the existing
experimental data to lower the RMSD.
This allows us  to make
reliable predictions for the ongoing experiments on ultracold
strontium molecules \cite{Zelevinsky:prive}. It turned out that
by slightly changing the $A_0$ and $C_{12}$ parameters by 0.03\% and 0.19\%, respectively, in the
analytical expression for the A$^1\Sigma_u^+$ potential,
Eq. (\ref{fitex}),  we could reduce the RMSD for $J'=1$ to 0.64 cm$^{-1}$.
With the new values of the $A_0$ and $C_{12}$ parameters, 
the root-mean-square-deviation  of our results for $J'\le 50$, as
compared to the raw data of Tiemann and collaborators 
(see the supplementary material of Ref. \cite{Tiemann:11})
is 4.5 cm$^{-1}$.
We must admit, however, that the present fit is not perfect for very high
values of $J'$. Fortunately, these values of the rotational quantum
number $J'$ are not of interest for the ultracold experiments such as
photoassociation spectroscopy \cite{ZelevinskyPRL06} or precision
measurements leading to the 
determination of the time variation of the electron to proton mass ratio
\cite{ZelevinskyPRL08,ZelevinskyPRA}. To better reproduce the levels
near the ${\rm ^1S_0 + {^3P_1}}$ threshold \cite{ZelevinskyPRL06}, 
we have  adjusted  the spin-orbit coupling between
the c$^3\Pi_u$ and B$^1\Sigma_u^+$ states, Eq. (\ref{xi2}). Specifically, 
we varied the parameters $X^{\infty}$, $B^X$, and $X_3$ in the fit of $\xi_2(R)$.
The adjustments
introduced to our {\em ab initio} data are 
illustrated in Fig. \ref{fig5}. Quantitatively, the 
adjusted and {\em ab initio}  potentials for the A$^1\Sigma_u^+$ state
differ by 290 cm$^{-1}$ in the repulsive region, at $R=6.5$ bohr, by 
-2.6 cm$^{-1}$ at the minimum, $R=7.54$ bohr, and by  0.07 cm$^{-1}$
in the long range, at $R=20$ bohr. All these differences represent
at most 2.5\% of the original {\em ab initio} potential. Such an accuracy
of the {\em ab initio} calculations for a system of this size would be
purely accidental.  
For the spin-orbit term  $\xi_2(R)$, the adjustment results in a change by
nearly 50\% in the repulsive region and less than 0.3\% in the asymptotic
value. Here, the shifts were relatively more important than
in the case of the  A$^1\Sigma_u^+$ potential, and this can be attributed to the lower
accuracy of the results from the MRCI method, which was employed for 
the calculation of the spin-orbit coupling. It should be stressed, however,
that only the long-range value of the $\xi_2(R)$ coupling has some  significance
for the described dynamics of the Sr$_2$ molecule, as it affects positions of the most weakly 
bound levels below the ${\rm ^1S_0 + {^3P_1}}$ threshold. 
Note that the parameters
reported in Tables \ref{tab1} and \ref{tab1a} are those adjusted to the
experimental data for $J'=1$.

\begin{figure}[t!]
\begin{center}
\vspace{-0.6cm}
\includegraphics[angle=-90,width=0.9\linewidth]{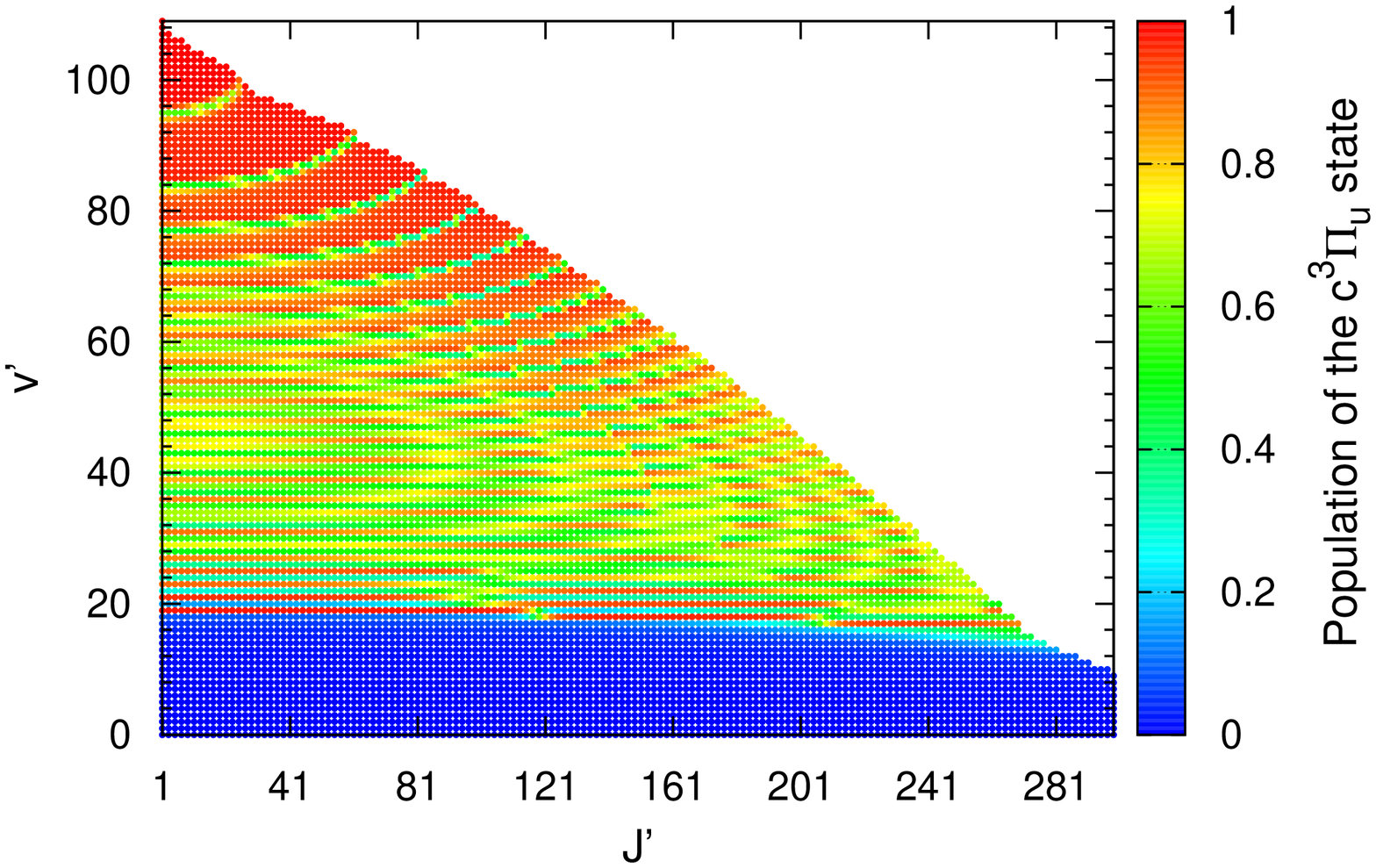}\vspace{-0.8cm}\\
\includegraphics[angle=-90,width=0.9\linewidth]{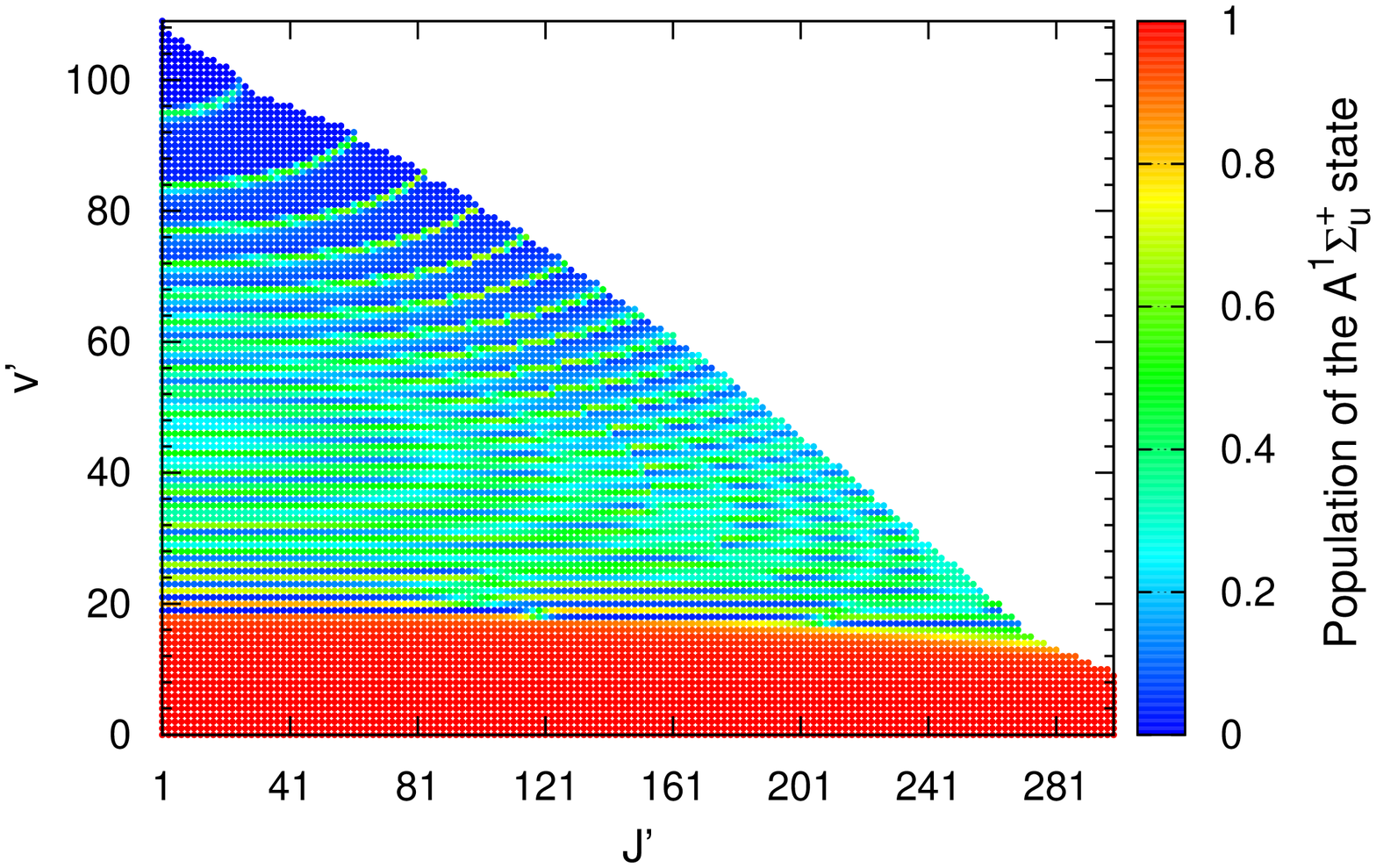}
\end{center}
\caption{
Population of the c$^3\Pi_u$ and A$^1\Sigma_u^+$ components
of the $0_u^+$ rovibrational levels lying below the $^1{\rm S_0 + {^3P_1}}$ asymptote.
}
\label{fig6}
\end{figure}

From Figs. \ref{fig1} and \ref{fig3} we expect that some of the
rovibrational levels can  be assigned to a single state, A$^1\Sigma_u^+$ or
c$^3\Pi_u$, while some of them will show a strongly mixed
singlet/triplet character. This is indeed the case, as is nicely
illustrated by Fig. \ref{fig6}. The contour plots of 
the A$^1\Sigma_u^+$ and c$^3\Pi_u$ state populations  
clearly show that for lower values of $J'$ all levels with
$v'\le 18$ are predominantly singlet levels of the A$^1\Sigma_u^+$
state. By contrast, the $v'=19$ level is the first level that
can fully be assigned to the c$^3\Pi_u$ state. This is also
clear by comparing the energies of this particular level in the coupled
model and in the Born-Oppenheimer approximation: 1363 cm$^{-1}$
vs. 1398 cm$^{-1}$. At higher values of $v'$ the situation becomes
quite erratic and most of the levels are of strongly mixed
singlet/triplet character. Only at high values of $v'$ the levels
can again be assigned, this time to the c$^3\Pi_u$ state, 
although some regions of strong mixing are observed,
e.g., near $v'=84$ or $v'=95$. Such a mixing in the rovibrational
levels close to the dissociation limit is very important for
the ongoing photoassociation experiments of ultracold strontium
atoms and is discussed in detail in Ref.
\cite{Skomorowski:12a}.
\begin{figure}[b!]
\begin{center}
 \includegraphics[angle=0,width=0.9\linewidth]{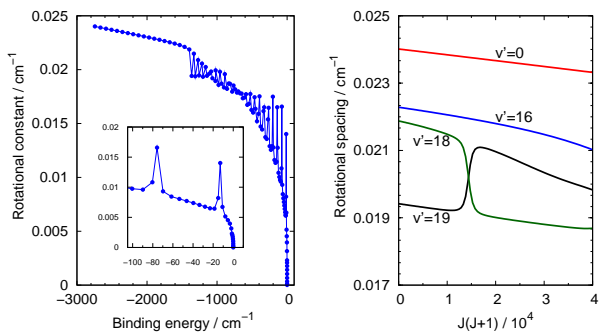}
\end{center}
\caption{
Rotational constant of the $0_u^+$ rovibrational levels for $J=1'$ (left-hand panel)
and rotational spacing, defined by $\Delta E_{vJ}/(4J-2)$
(right-hand panel).
}
\label{fig7}
\end{figure}

More detailed information on the strongly mixed levels
is obtained by examing the rotational constants $B_{v'}$
as function of the binding energy of the rovibrational level
$|v',J'\rangle$. This is illustrated in the left-hand panel of
Fig. \ref{fig7},  showing  the first 19
levels from $v'=0$ to 18 not to be perturbed, with 
a uniform nearly linear dependence of the rotational constant 
on the binding energy.
Starting from $v'=18$ the linear dependence is clearly
broken, and the curve is characterized by irregular behaviour. 
However, it is still possible to distinguish levels which are almost
purely triplet. 
The rotational spacings as functions of $J'(J'+1)$
reported in the right-hand panel of Fig. \ref{fig7} give 
information about the perturbed levels. Indeed, the rotational
spacing for $v'=0$ shows a purely linear character as a function
of $J'(J'+1)$. For $v'=16$ this linear dependence is broken for
high values of $J'$, but the nonlinear character  is due
to the centrifugal distortion, and not to the triplet
perturbations only. Starting from $v'=18$,
the rovibrational levels can have very different character (purely triplet or
singlet, strongly mixed), depending on the value of $J'$.
This is clearly illustrated by the sudden changes in the slope of
the linear dependence of the rotational spacing on $J'(J'+1)$.
\begin{figure}[t!]
\begin{center}
\includegraphics[angle=-90,width=0.9\linewidth]{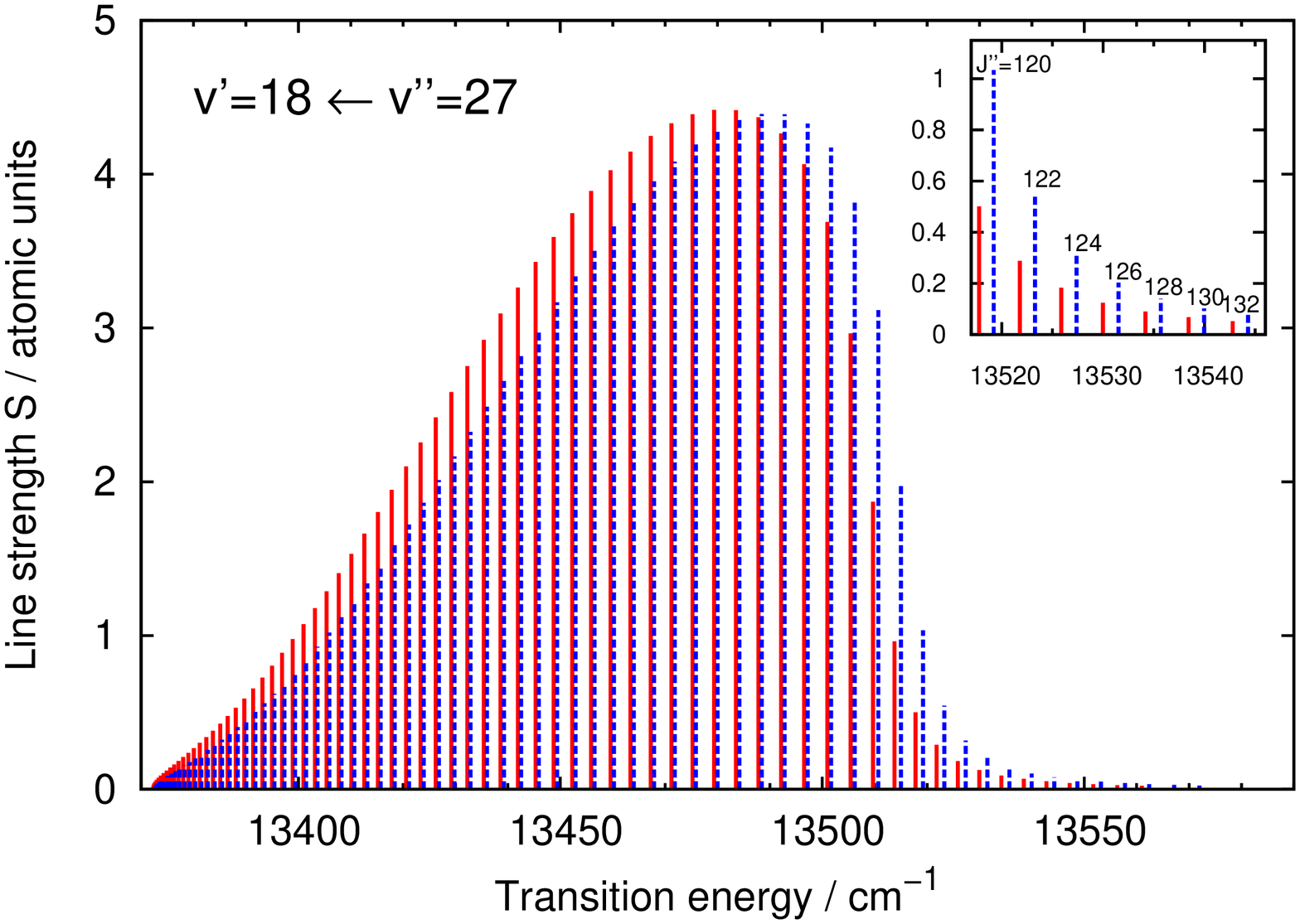}\vspace{-0.0cm}\\
\includegraphics[angle=-90,width=0.9\linewidth]{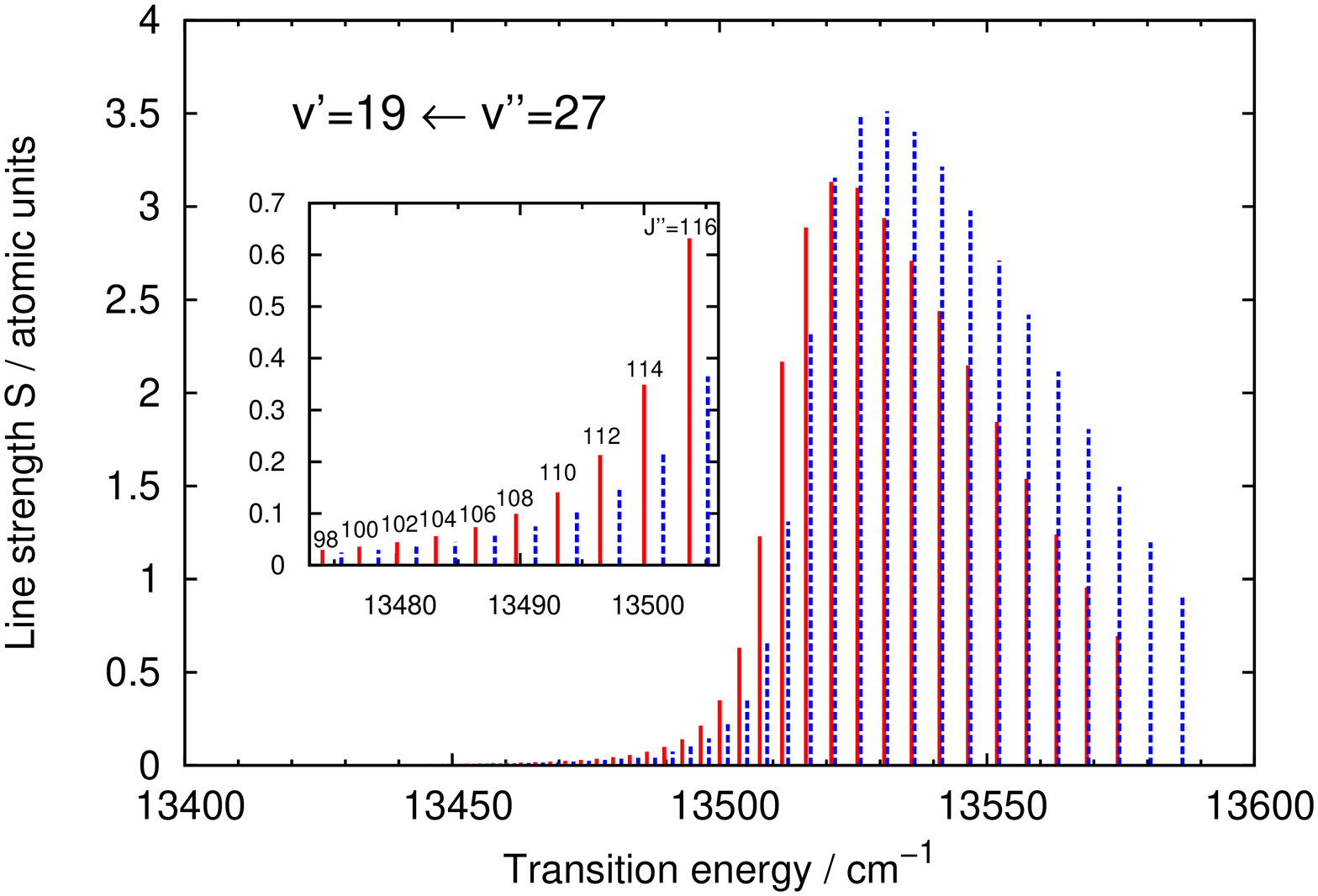}\end{center}
\caption{
Line strengths corresponding to the $P$ (red solid) and $R$ (blue dashed) branches
in the band $\nu '=18 \leftarrow \nu ''=27$ (upper panel) and
$\nu '=19 \leftarrow \nu ''=27$ (lower panel).  The spectral lines in the figures
are labeled by the rotational quantum number $J''$ of the initial state.
}
\label{fig8}
\end{figure}

Figure \ref{fig8} reports the line strengths for the
transitions $v'=18 \leftarrow v''=27$ and $v'=19 \leftarrow v''=27$
for different values of the rotational quantum number $J''$ of the initial
state. The level $v''=27$ was chosen since it plays a central role in
proposed experiments to determine a time variation of the
electron to proton mass ratio~\cite{ZelevinskyPRL08,ZelevinskyPRA}.
Inspection of Fig.~\ref{fig8} reveals that the spectral line strengths
for the $v'=18 \leftarrow v''=27$ transitions
show a typical dependence on the transition frequency, with a slow increase
as a function of $J''$, and a relatively steep and fast decrease for large
transition frequencies. By
contrast, the line strengths for the $v'=19 \leftarrow v''=27$
transitions show a completely different dependence on the transition
frequency. First we observe a very slow increase of the line strength
with the increasing $J''$, see the insert and note the difference in the
intensity scale, and a very slow decrease at large values of the transition
frequency. Such a behaviour is again a signature of strong perturbations from
the c$^3\Pi_u$ state that should be observable in experiment. 

\subsection{Feshbach resonances above the $^1{\rm S_0}+^3{\rm P}_1$ asymptote}
\label{sub6}

The rovibrational states that could be assigned to the
(2)$0_u^+$ state lying above the $^1{\rm S_0 + {^3P_1}}$ dissociation
limit are embedded in the continuum of the (1)$0_u^+$ state, and thus are
Feshbach resonances, i.e., quasi-bound states with finite lifetimes. Two
decay mechanisms are possible depending on the strength of the coupling
with the continuum. The resonances can decay to the $^1{\rm S_0 + {^3P_1}}$
continuum, i.e., dissociate into atoms (predissociation), or, if the
lifetime is long enough, decay to bound or continuum states of the
electronic ground state (spontaneous emission). 
Our computed positions, widths, and lifetimes of a few low lying 
Feshbach resonances for $J'=1$ are reported in Table \ref{tab4}.
The width of the resonances
varies quite strongly, between almost 1 cm$^{-1}$ for the very broad
one to $\approx$0.001 cm$^{-1}$ for the very narrow one. The
corresponding lifetimes also vary considerably, from the picosecond
to the nanosecond scale. This implies that some of the resonances
should be observable 
in high-resolution Fourier transform spectroscopic experiments.
\begin{table}[b!]
\caption{
Positions (relative to the $^1{\rm S_0 + {^3P_1}}$ threshold), widths, and lifetimes
of the Feshbach resonances of $^{88}$Sr$_2$ above the $^1{\rm S_0 + {^3P_1}}$ dissociation
limit.
\label{tab4}}
\vskip 5ex
\begin{tabular}{cccc}
\hline\hline
\clc{$v'$} & \clc{\hspace{0.3cm}  $E_r$/cm$^{-1}$ \hspace{0.3cm}}  &  \clc{\hspace{0.3cm} $\Gamma$/cm$^{-1}$ \hspace{0.3cm}} &   \clc{\hspace{0.3cm} $\tau$/ns \hspace{0.3cm}}  \\
    \hline
110  & { 49.8}     &   0.8761&  0.0037  \\
111  & 112.2     &   0.4658&  0.0067  \\
112  & 174.4     &   0.1924&  0.0162  \\
113  & 236.3     &   0.0410&  0.0759  \\
114  & 297.9     &   0.0008&  3.8821  \\
115  & 359.2     &   0.0638&  0.0488  \\
\hline\hline
\end{tabular}
\end{table}

\begin{figure}[t!]
\begin{center}
 \includegraphics[angle=0,width=0.9\linewidth]{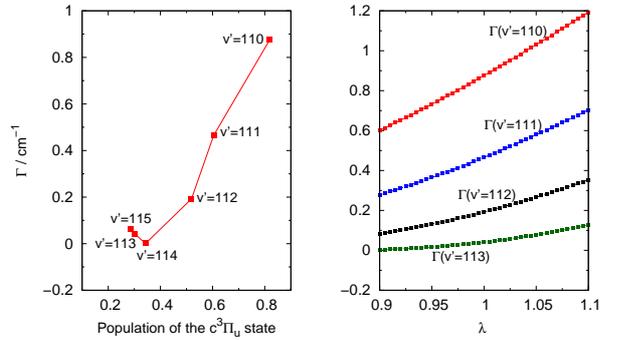}
\end{center}
\caption{
Widths of the Feshbach resonances as a function of the population
of the c$^3\Pi_u$ state and scaling parameter $\lambda$ of the
spin-orbit term  $\langle {\rm c}^3\Pi_u |H_{\rm SO}|{\rm A}^1\Sigma_u^+ \rangle$.
}
\label{fig9}
\end{figure}

In Table \ref{tab4} we have numbered the resonant states by the
vibrational quantum number $v'$ corresponding to bound vibrational
states above the $^1{\rm S_0 + {^3P_1}}$  dissociation limit
that result from the diagonalization of the $\widehat{{\Bbb H}}^{0_u^+}$
Hamiltonian. For $J'=1$ we have found 110 bound levels of
the $0^+_u$ symmetry which lie below the $^1{\rm S_0 + {^3P_1}}$ asymptote, 
assigning them vibrational quantum numbers ranging $v'=0$ to $v'=109$.
Thus the first quasi-bound level reported in Table \ref{tab4} 
located above the $^1{\rm S_0 + {^3P_1}}$ threshold  
has  the vibrational quantum number $v'=110$. 
Only eigenvalues that were stable with respect to the grid
size were selected, and the exact positions and lifetimes were
determined with the complex absorbing potential of Eq. (\ref{CAP}).
It turns out that the width/lifetime is a very sensitive function
of the strength of the spin-orbit coupling. To illustrate this point, 
we report the widths of the resonant states as a function of the
triplet state population of the rovibrational wave function in the
left-hand  panel of Fig. \ref{fig9}.  
The narrow resonances are found to be dominated by the singlet
component of the wave function, for which 
the coupling through the spin-orbit interaction is weak. By
contrast, those resonances that are almost purely triplet states are very
broad with short, picosecond lifetimes. 
This is easily rationalized in terms of a state
with a predominantly singlet character 
decaying more slowly to the triplet continuum than
an (almost) purely triplet state. Visibly, the spin-orbit coupling is responsible
for  the predissociation decay of the Feshbach resonances, i.e., 
a smaller spin-orbit interaction favours longer-lasting 
quasi-bound states, cf. $v'=114$ reported in Table \ref{tab4}.  

It was shown in Ref. \cite{Moszynski:94} that the widths of the
resonances are very sensitive to the quality of the {\em ab initio}
data. Since the existence of low-lying Feshbach resonances is mostly
due to the spin-orbit coupling  $\xi_1(R)$, between the ${\rm
  c}^3\Pi_u $ and ${\rm A}^1\Sigma_u^+$ states, we have checked how
scaling this particular term by a parameter $\lambda$ ranging from 0.9
to 1.1,  $\lambda \cdot  \xi_1(R)$, affects the results.
This is illustrated in the right-hand  panel of Fig. \ref{fig9}. Inspection
of this figure shows  the width to be indeed a rather sensitive function
of the coupling, and a 10\% change in the coupling may result in a
change of the width by a factor of two. However, since scaling
within $\pm 10\%$ does not change the order of magnitude of  the
lifetimes, 
our conclusions concerning a possible experimental observation
of the Feshbach resonances in the A$^1\Sigma_u^+$, c$^3\Pi_u$, and
a$^3\Sigma_u^+$ manifold remain valid.

\section{Summary and conclusions}
\label{sec4}
We have reported a theoretical study of interatomic interactions
and spectroscopy of the strontium dimer in the A$^1\Sigma_u^+$, c$^3\Pi_u$,
and a$^3\Sigma_u^+$ manifold of electronic states. 
The spectroscopic characteristics of the potential energy curves for the 
nonrelativistic A$^1\Sigma_u^+$, c$^3\Pi_u$, and a$^3\Sigma_u^+$ states
of Sr$_2$ and of the relativistic states of $0_u^+$ and $1_u$ symmetry
deviates significantly from most of the previous {\em ab initio} results
\cite{Czuchaj:03,Koto:08}. This is particularly true for the 
optically active A$^1\Sigma_u^+$ state and its Hund case $(c)$ analog, 
(1)$0_u^+$.
By contrast, the present spectroscopic characteristics of the (1)$0_u^+$
state are in a very good agreement with the experimental values deduced
from the high-resolution Fourier transform spectroscopic data \cite{Tiemann:11}.
The rovibrational energy levels corresponding to the spin-orbit coupled
(1)$0_u^+$ state dissociating into ${\rm ^1S_0+ {^3P}_1}$ atoms lying below the 
avoided crossing with the (2)$0_u^+$ state dissociating into ${\rm ^1S_0+{^1}D_2}$ 
atoms are almost unperturbed, and the corresponding energies are very close
to the energies obtained in the Born-Oppenheimer approximation. 
The rovibrational levels of the (1)$0_u^+$ state lying above the avoided
crossing with the (2)$0_u^+$ state are all heavily perturbed by the
rovibrational states of the c state. These perturbations
are exclusively due to the spin-orbit interaction.
In all cases, the nonadiabatic effects due to the Coriolis couplings 
were shown to be negligible, with the exception of a few least bound levels.
We have located several  quasi-bound Feshbach resonances lying above the
${\rm ^1S_0 + {^3P_1}}$ dissociation limit. Their lifetimes
suggest that they should be observable  in high-resolution spectroscopic
experiments. 

Overall,  our results reproduce (semi)quantitatively the 
experimental data observed thus far \cite{Tiemann:11}. 
Our spectroscopic predictions for on-going and future experiments concerning the
photoassociation of ultracold strontium atoms \cite{ZelevinskyPRL06,Zelevinsky:prive}
and precision measurements of the time variation of the electron to proton
mass ratio \cite{ZelevinskyPRL08,ZelevinskyPRA} are  reported elsewhere \cite{Skomorowski:12a}.

\section*{Acknowledgments}
We would like to thank Eberhard Tiemann, Paul Julienne, Svetlana Kotochigova and  Tanya
Zelevinsky for many useful discussions.
Financial support from the Polish Ministry of Science and
Higher Education through the project
N N204 215539 and from the Deutsche Forschungsgemeinschaft through the
Emmy Noether programme are gratefully acknowledged.


%
\end{document}